\documentclass[epj-spec]{svjour}
\usepackage{graphicx}
\begin{document}
\title{Properties of quark gluon plasma from lattice calculations}
\author{P\'eter Petreczky\inst{1}\fnmsep\thanks{\email{petreczk@bnl.gov}}  }
\institute{Physics Department and RIKEN-BNL Research Center, Brookhaven National Laboratory, Upton, NY 11973 USA}
\abstract{
I discuss lattice QCD calculations of the properties of strongly interacting matter at finite temperature, including the
determination of the transition temperature $T_c$, equation of state, different static screening
lengths and quarkonium spectral functions.
The lattice data suggest that at temperatures above $2.0T_c$ many properties of the quark gluon plasma can
be understood using weak coupling approach, although non-perturbative effects due to static magnetic fields
are significant in some quantities.
} 
%
\maketitle  
\section{Introduction}
\label{intro}

One expects that at sufficiently high
temperatures and densities the strongly interacting matter undergoes a
transition to a new state, where quarks and gluons are no longer confined in
hadrons, and which is therefore often referred to as a deconfined phase
or Quark Gluon Plasma (QGP).
The main goal of heavy ion experiments is to
create such form of matter and study its properties.
We would like to know at which temperature the transition takes place and what
is the nature of the transition as well the properties of the deconfined
phase, equation of state, static screening lengths, transport properties etc.
Lattice QCD can provide first principle calculation of the transition temperature,
equation of state and static screening lengths (see Ref. \cite{sewm06,lat06} for recent reviews ). 
Calculation of transport coefficients remains an open challenge for lattice QCD 
(see discussion in Refs. \cite{aarts,derek}).
 
 One of the most interesting question for the lattice
is the question about the nature of the finite temperature transition
and the value of the temperature $T_c$ where it takes place.
For very heavy quarks we have a 1st order deconfining transition.
In the
case of QCD with three  flavors of quarks we expect a 1st order
chiral transition for sufficiently small quark masses.
In other cases there is no true phase transition but just a rapid
crossover.
Lattice simulations of 3 flavor QCD with improved staggered quarks (p4) using
$N_{\tau}=4$ lattices indicate that
the transition is first order only for very small quark masses,
corresponding to pseudo-scalar meson masses of about $60$ MeV
\cite{karschlat03}.
A recent study of the transition using effective models
of QCD  resulted in a similar estimate for the boundary in the quark mass
plane, where the transition is 1st order \cite{szepzs}.
This makes it unlikely that for the interesting case of one heavier strange
quark and two light $u,d$ quarks, corresponding to $140$ MeV pion, the
transition is 1st order. However, calculations with unimproved staggered
quarks suggest that the transition is 1st order for pseudo-scalar
meson mass of about $300$ MeV \cite{norman}.
Thus the effect of the improvement is
significant and we may expect that the improvement of flavor symmetry,
which is broken in the staggered formulation, is very important.
But even when using improved staggered fermions it is necessary to do the calculations at several
lattice spacings in order to establish the continuum limit.
Recently,  extensive calculations have been done to clarify the nature
of the transition in the 2+1 flavor QCD for physical quark masses using
$N_t=4,~6,~8$ and $10$ lattices.
These calculations were done using the
so-called $stout$ improved staggered fermion formulations which improves
the flavor symmetry of staggered fermions but not the rotational symmetry. 
The result of this study was
that the transition is not a true phase transition but only a rapid
crossover \cite{nature}.
Even-though there is no true phase transition in  QCD thermodynamic observables change rapidly in a small
temperature interval and the value of
the transition temperature plays an important role.
The flavor and quark mass dependence of
many thermodynamic quantities is largely determined by the flavor and
quark mass dependence of $T_c$. For example, the pressure normalized by
its ideal gas value for pure gauge theory, 2 flavor, 2+1 flavor and 3 flavor
QCD shows almost universal behavior as function of $T/T_c$ \cite{cargese}.
\begin{figure}
\includegraphics[width=7cm]{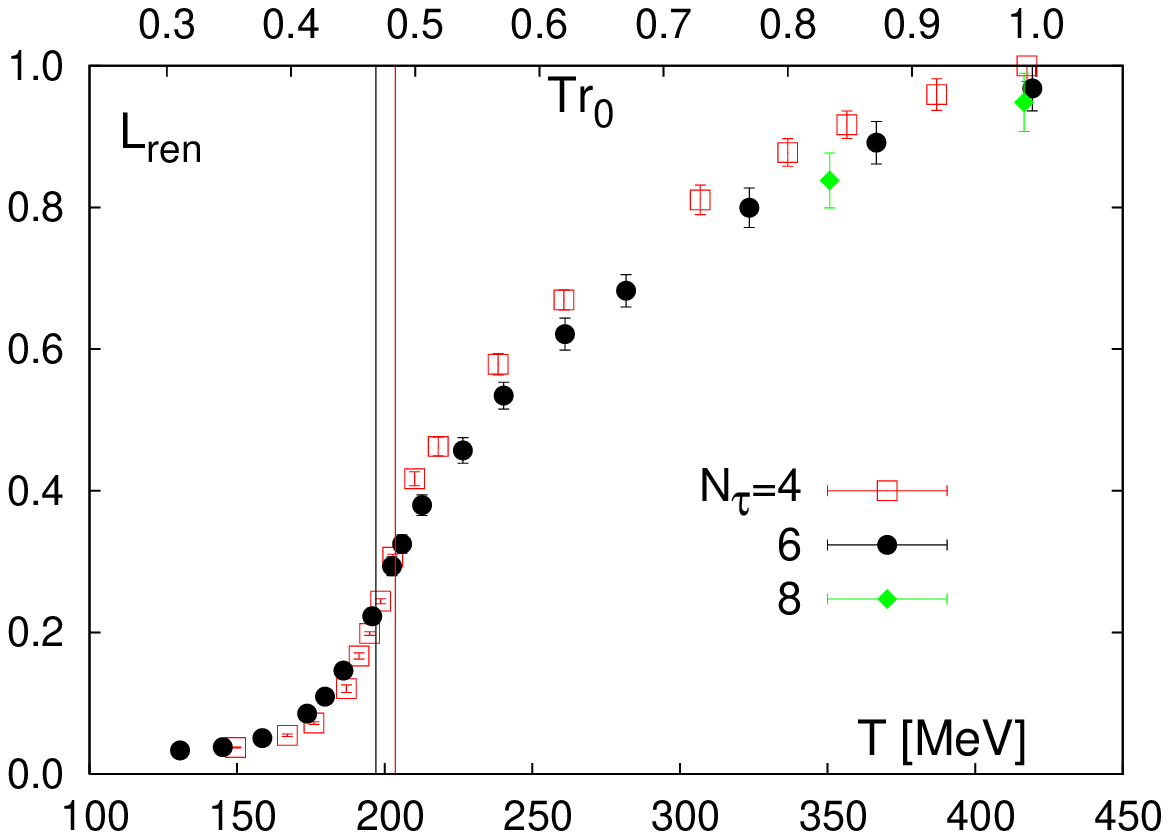}
\includegraphics[width=7cm]{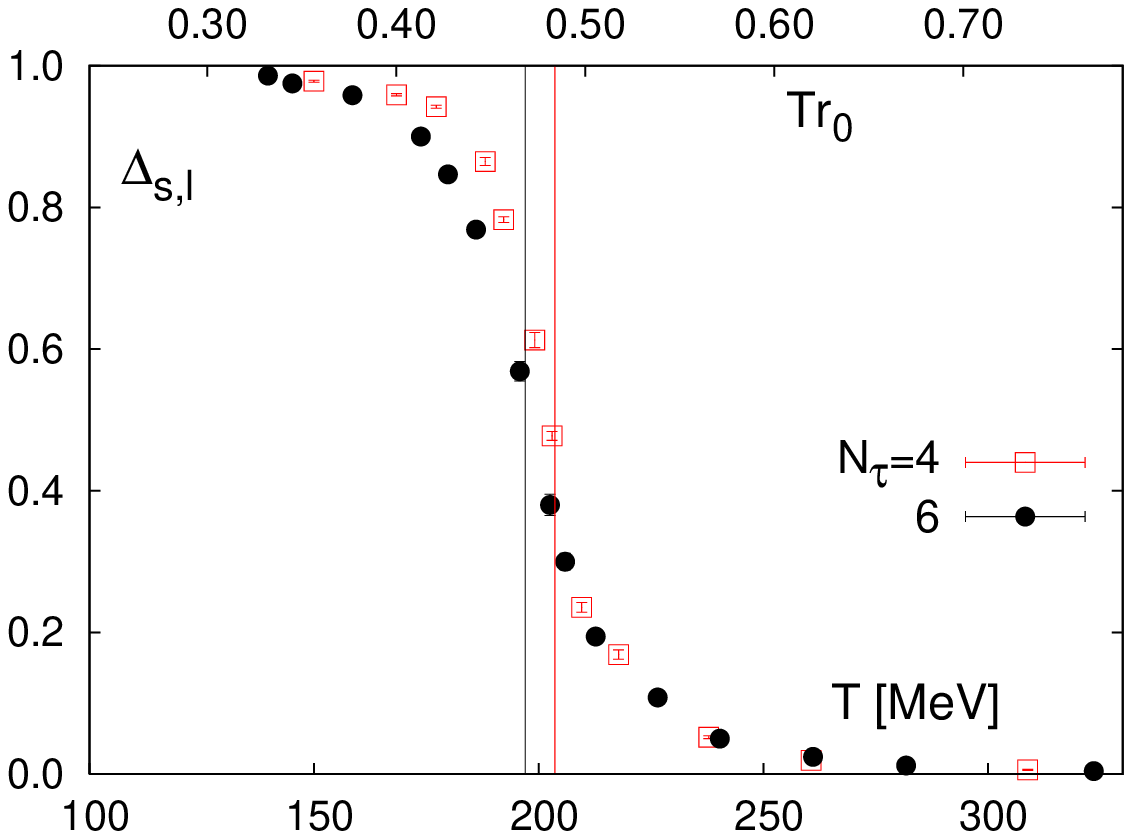}
\caption{The renormalized Polyakov loop $L_{ren}(T)$ (left) and the chiral condensate normalized to the zero
temperature chiral condensate (right) as function of the temperature calculated at $m_l=0.1m_s$ \cite{eos}.}
\label{fig:order}
\end{figure}
The chiral condensate $\langle \bar \psi \psi \rangle$ and the expectation value of the Polyakov loop $\langle L \rangle$
are order parameters in the limit of vanishing and infinite quark masses respectively. However, also for finite values of
the quark masses they show a rapid change in vicinity of the transition temperature. In Figure \ref{fig:order} I show
the chiral condensate and the Polaykov loop as function of the temperature calculated for the $p4$ action and light 
quark mass $m_l=0.1m_s$, with $m_s$ being the physical strange quark mass. 
The expectation value of the Polyakov loop vanishes in the continuum limit unless it
is renormalized ( see e.g. \cite{pisarski}). The renormalized Polyakov loop can be defined 
through the free energy of a static quark anti-quark free energy, $F(r,T)$ 
normalized to the zero temperature potential at short distances \cite{okacz02}. 
\begin{equation}
L(T)=\exp(-\frac{F_{\infty}(T)}{2T}),
\end{equation}
where $F_{\infty}(T)=\lim_{r \rightarrow \infty} F(r,T)$. Therefore what is shown in Figure \ref{fig:order} is
the renormalized Polyakov loop \cite{eos}.  In fact it is not necessary to calculate the
free energy for the renormalization of the Polyakov loop. It is sufficient to calculate the zero
temperature potential at each lattice spacing where finite temperature calculations have been 
performed \cite{fodorplb}. This method  gives more precise result for the renormalized Polyakov
loop then the one used in Ref. \cite{okacz02}., where the free energy has been normalized to the $T=0$
potential at the shortest distance.
The difference in the value of the renormalized Polyakov loop in Figure \ref{fig:order} and in Ref. \cite{fodorplb}
is due to the different normalization of the zero temperature potential.  
We see that the chiral condensate and the renormalized Polyakov loop show rapid change at $T_c$ suggesting that
the chiral and the deconfinement transitions happen at the same temperature. 

To determine the value of the transition temperature
and to study the interplay between the chiral and the deconfinement transition one usually calculates the disconnected part of the
chiral susceptibility and the Polyakov loop susceptibility defined as 
\begin{equation}
\frac{\chi_{\bar \psi \psi}}{T^2}=N_{\sigma}^3 ( \langle (\bar \psi \psi)^2\rangle - \langle \bar \psi \psi \rangle^2),~~~~~
\frac{\chi_{L}}{T^2}=N_{\sigma}^3 ( \langle L^2\rangle - \langle L \rangle^2)
\end{equation}
as function of the of the bare gauge coupling $\beta=6/g^2$. Here $N_\sigma$ is the spatial size of the lattice. The susceptibilities
have a peak at some pseudo-critical coupling $\beta_c$. 
The chiral and the Polyakov loop susceptibility have been studied using lattice with temporal extent $N_{\tau}=4$ and 
$N_{\tau}=6$ and several values of the light quark masses $m_l=0.05m_s,~0.1m_s,~0.2m_s$ and $0.4m_s$ \cite{us06}.
Note that the smallest value of $m_l$ correspond to pion masses of about $140$MeV. We find that within accuracy of the 
calculations pseudo-critical couplings $\beta_c$ determined from the disconnected part of the chiral susceptibility and
the Polyakov loop susceptibility coincide. This again shows that the chiral and the deconfinement transition happen at
the same temperature.

To determine the transition temperature we have to calculate the lattice spacing in terms of some physical quantity.
In the past the string tension has been used to set the lattice spacing. A more accurate determination of the lattice spacing
is provided by the so-called Sommer scale $r_0$ defined from the static quark anti-quark potential as 
$r^2 \frac{d V(r)}{d r}|_{r=r_0}=1.65.$
Analysis of the quarkonium spectroscopy on the lattice lead to  the value $r_0=0.469(7)$fm \cite{gray}. In Figure \ref{fig:tcr0}
I show the transition temperature in units of $r_0$ for different quark masses \cite{us06} and two different lattice spacings, corresponding 
to $N_{\tau}=4$ and $N_{\tau}=6$ lattices. Note that the value of $T_c$ calculated at two different lattice
spacings are clearly different. The thin error-bars in Figure \ref{fig:tcr0}  represent the
error in the determination of the lattice spacing $a$, i.e.
the error in $r_0/a$. There is also an error in the determination of the
pseudo-critical coupling constant $\beta_c=6/g^2$. The combined error is shown in
Figure \ref{fig:tcr0} as a thick error-bar. For $N_t=4$ calculations the error is dominated
by the error in lattice spacing, while for $N_t=6$ it is dominated by the
error in $\beta_c$. With the data on $r_0 T_c$ a chiral and continuum
extrapolation has been attempted using the most simple Ansatz
$r_0 T_c(m_{\pi},N_t)=r_0 T_c|_{cont}^{chiral}+ A (r_0 m_{\pi})^d + B/N_t^2$.
From this extrapolation on gets the continuum value $T_c r_0=0.457(7)(+8)(-2)$
for the physical pion mass $m_{\pi} r_0=0.321$ {\cite{us06}.
The central value was obtained using
$d=1.08$ expected
from $O(4)$ scaling. To test the sensitivity to the chiral extrapolations
$d=2$ and $1$ have also been used.
The resulting uncertainty is shown
as second and third error.
Using the
best know value of $r_0=0.469(7)$fm we obtain
$T_c=192(7)(4)$ MeV
which is higher than the most
of the previous values. It is also significantly higher than the
chemical freezout temperature at RHIC \cite{starwhite}. Note that the large value of the transition temperature is mostly due
to the large value of the string tension which is related to the Sommer scale as $r_0 \sqrt{\sigma}=1.114(4)$ \cite{milc01}. 
Using the above value of $r_0$ we get $\sqrt{\sigma}=468$ MeV which is more than $10\%$
larger than the value $\sqrt{\sigma}=420$ MeV which was used in Ref. \cite{karsch01} and led to $T_c=173(8)$MeV. Recently the transition temperature has been
determined using the so-called $stout$ staggered action and $N_t=4,~6,~8$ and $10$
\cite{fodorplb}. The deconfinement temperature has been found to be 
$176(3)(4)$ MeV \cite{fodorplb}. The central 
value is considerably smaller than the
one obtained with $p4$ action but taking into account the errors the deviation
is not very significant. Calculations on $N_{\tau}=8$ lattices 
with $p4$ action are needed to clarify this issue. 
The authors of Ref. \cite{fodorplb} use a different definition of the chiral susceptibility which resulted
in the chiral transition temperature of $T_{chiral}=151(3)(3)$MeV. Using the
chiral susceptibility defined above would result in a larger value of the
transition temperature.  
\begin{figure}
\includegraphics[width=7cm] {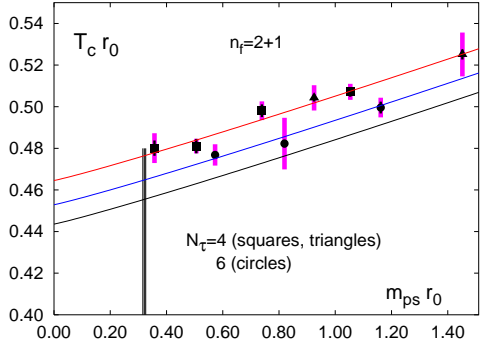}
\caption{The transition temperature in units of the $r_0 T_c$ from Ref. \cite{us06}
as function of the pion mass.}
\label{fig:tcr0}
\end{figure}

Lattice calculations of equation of state were started some twenty years ago. In the case of QCD without dynamical
quarks the problem has been solved, i.e. the equation of state has been calculated in the continuum limit \cite{boyd96}.
At temperatures of about $4T_c$ the deviation from the ideal gas value is only about $15\%$ suggesting that quark gluon
plasma at this temperate is weakly interacting. Perturbative expansion of the pressure, however, showed very poor
convergence at this temperature \cite{arnold}. Only through the use of new re-summed perturbative techniques it was possible
to get agreement with the lattice data \cite{scpt97,braaten,blaizot}. To get a reliable calculation of the pressure and the energy 
density improved action have to be used \cite{heller99,karsch00}. Very recently calculations with the so-called $asqtad$ and $p4$
action have been done on lattices with temporal effects $N_{\tau}=4$ and $6$ \cite{milc06,eos}.  In Figure \ref{fig:e-3p}
the interaction measure $\epsilon-3p$ as well as the pressure and the energy density are shown as function of the temperature for the $p4$ action. 
Calculations performed for
$N_{\tau}=4 $ and $N_{\tau}=6$ give similar results, e.g. the difference between them is at most $7\%$ for the pressure. The difference 
in the height of the interaction measure close to $T_c$ is only due to the non-perturbative features of the beta function at large lattice
gauge coupling \cite{eos}.
Therefore cutoff effects are more or less under control in these calculations. Furthermore, there is
a good agreement between $p4$ and $aqstad$ calculations \cite{milc06}. We see that close to $T_c$
the interaction measure is very large, which means that quark gluon plasma at this temperature is very far from the conformal limit, where $\epsilon=3 p$.
At high temperature the value of the interaction measure is consistent with the perturbative estimate.
The pressure is about $10\%$ below the ideal gas value and thus it is closer to the ideal gas limit as in the previous calculations \cite{karsch00}.
This is due to the fact that the previous calculations were
done at bare  quark masses fixed in units of the temperature $m_l/T=0.4$, while the new calculations have been performed for 
bare quark masses which correspond to fixed pion mass of about $220$ MeV and physical kaon mass. This conclusion is also supported by
the analysis of the equation of state in Ref. \cite{fodor_eos04}.
\begin{figure}
\includegraphics[width=7cm]{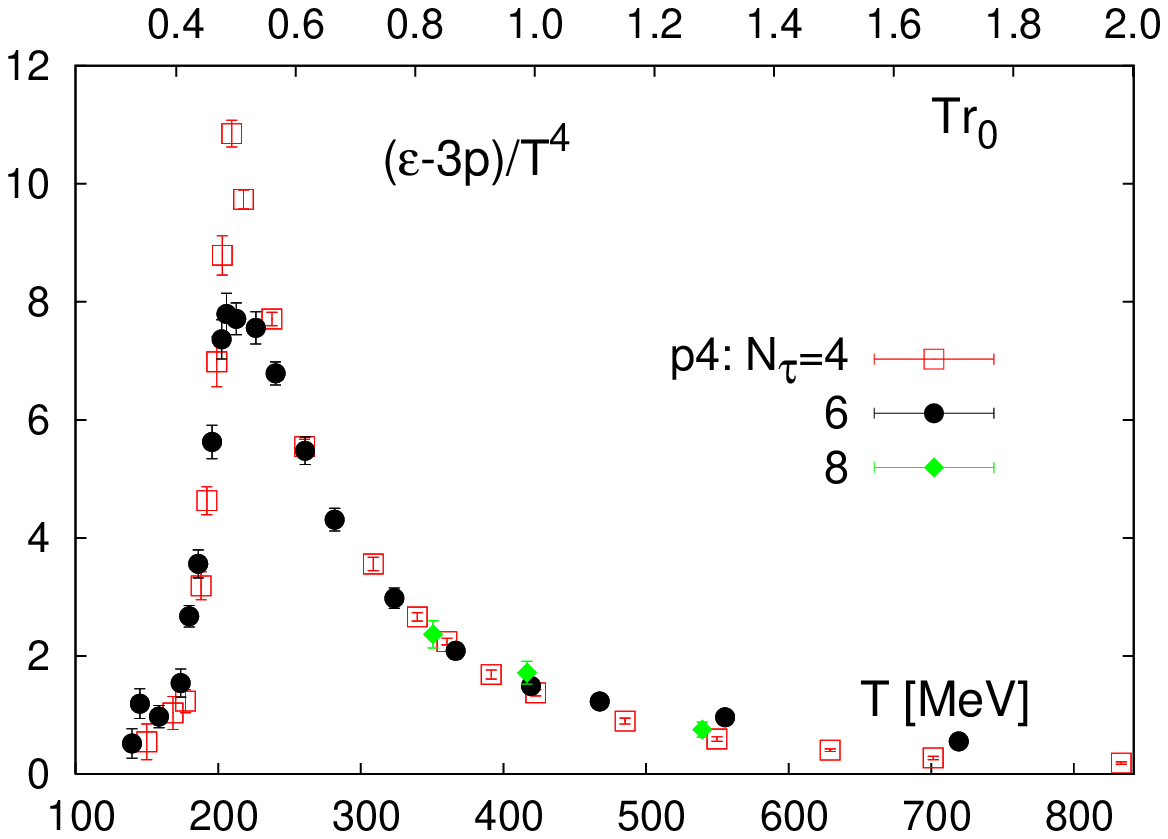}
\includegraphics[width=7cm]{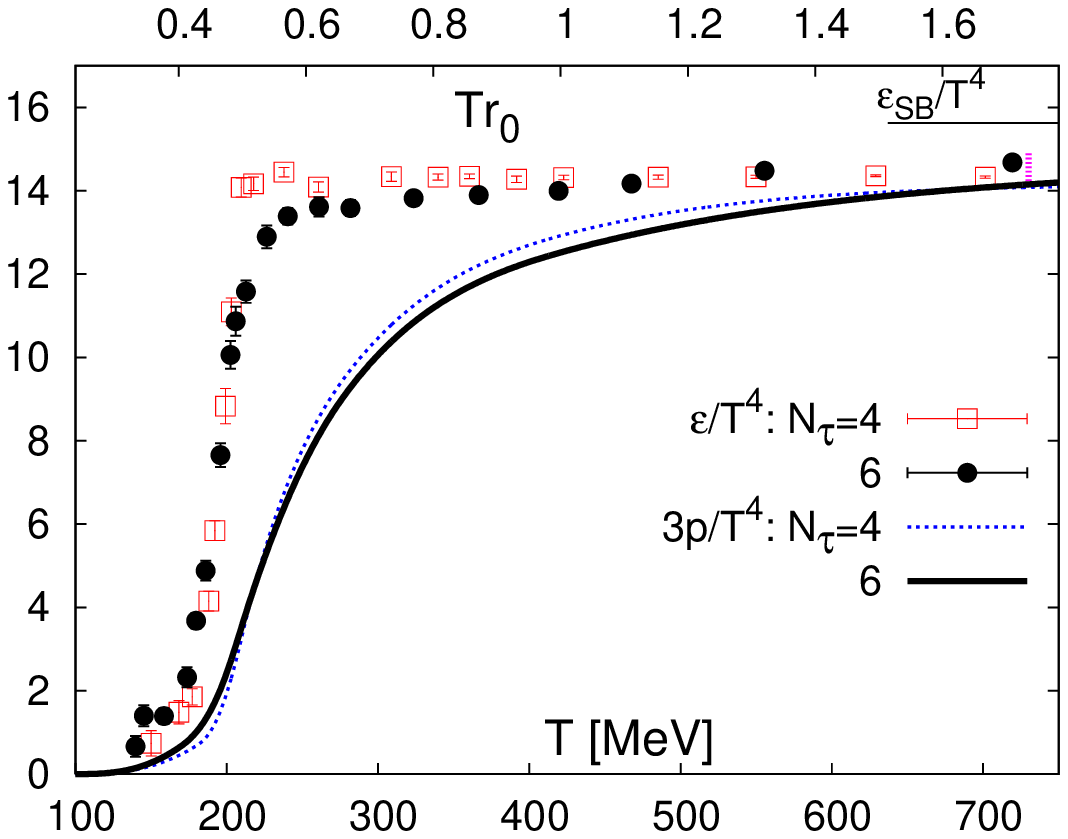}
\caption{ 
The interaction measure calculated (left) as well as  the pressure and energy density (right) for the p4 action \cite{eos}.}
\label{fig:e-3p}
\end{figure}

\section{Spatial correlation functions}
To get further insight into properties of the quark gluon plasma one can study  different spatial correlation functions.
One of the most prominent features of the quark gluon plasma is the presence of chromo-electric (Debye) screening.
The easiest way to study chromo-electric screening is to calculate the singlet free energy of static quark anti-quark pair (for a recent
review on this see Ref. \cite{mehard04}),  which is expressed in term of correlation function of temporal Wilson lines
\begin{equation}
\exp(-F_1(r,T)/T)={\rm Tr} \langle W(r) W^{\dagger}(0) \rangle.
\end{equation}
$L={\rm Tr}  W$ is the Polyakov loop.
Since the above correlator is not gauge invariant all calculations have been done in Coulomb gauge \cite{okacz02,kostya1,okacz04,okacz05,kostya_lat07}.
Alternatively one can insert spatial transporters between the two temporal Wilson lines which makes the correlator gauge invariant.
Calculations of such gauge invariant correlator have also been performed and lead to results which are very similar to those
obtained in Coulomb gauge \cite{okacz_lat01,okacz_tbp}.
The singlet free energy is also useful to study quarkonia binding at high temperatures 
\cite{digal01,wong,alberico,rapp,mocsy06,mocsy07}.  

In purely gluonic theory the free energy grows linearly with the separation between the heavy quark and 
anti-quark in the confined phase. In presence of dynamical quarks the free energy is saturated at some finite value 
at distances of about $1$ fm due to string breaking \cite{mehard04,kostya1,okacz04}. Above the deconfinement temperature the singlet free
energy is exponentially screened, at sufficiently large distances \cite{okacz04}, i.e.
\begin{equation}
F_1(r,T)=F_{\infty}(T)-\frac{4}{3}\frac{g^2(T)}{4 \pi r} \exp(-m_D(T) r).
\end{equation}
The inverse screening length or equivalently the Debye screening mass $m_D$ is proportional to the temperature. In leading order
of perturbation theory it is $m_D(T)=\sqrt{1+N_f/3} g(T) T$.
\begin{figure}
\includegraphics[width=8cm]{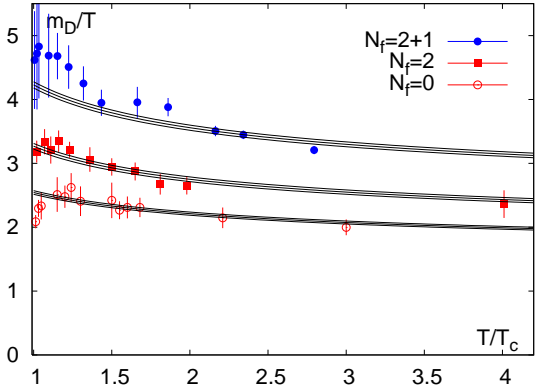}
\caption{
The Debye mass calculated in quenched QCD \cite{okacz04}, 2 flavor QCD \cite{okacz05} an in 2+1 flavor QCD. 
The lines show the leading order
fit together with its uncertainty.}
\label{fig:mD}
\end{figure}
Beyond leading order it is sensitive to the non-perturbative dynamics of the static chromo-magnetic fields. 
The Debye screening mass has been calculated in pure gauge theory ($N_f=0$) \cite{okacz04}, in 2 flavor QCD  ($N_f=2$) \cite{okacz05}
and very recently also in 2+1 flavor QCD \cite{kostya_lat07}
and is shown in Figure \ref{fig:mD} at different temperatures.  The temperature dependence of the lattice data have been fitted with the simple 
Ansatz motivated by the leading order result : $m_D(T)=A \sqrt{1+N_f/3} g(T) T$. Here $g(T)$ is the two loop running coupling constant in $\overline{MS}$ 
scheme with $\mu_{\overline{MS}}=2 \pi T$.
This simple form can fit the data very well and we get similar values of $A \simeq 1.4-1.5$ for different flavor content. 
Thus the temperature dependence as well as the flavor dependence of the Debye mass is given by perturbation
theory. We also see that non-perturbative effects due to static magnetic fields significantly effect the electric screening,
resulting in about $40-50\%$ corrections.  However, the non-perturbative correction  is the same in full QCD and pure gauge theory. 
Let us also note that in SU(2) gluodynamics the non-perturbative 
corrections to the Debye mass are approximately the same \cite{heller97,oevers98,cucchieri01}. 
This situation can be understood in terms of dimensionally reduced effective theory, where the effect of hard modes with momentum $p \sim \pi T$ 
is integrated out and which contain only static electric and magnetic fields \cite{kajantie97}. The validity of dimensional reduction has been tested
in a wide temperature range \cite{oevers98,cucchieri01}.

At zero temperature the static quark anti-quark potential is determined from the Wilson loops: $V(r)=-1/t \ln W(r,t),~t \rightarrow \infty$. At large
separation the Wilson loop obeys the area law $W(r,t) \sim \exp(-\sigma r t)$ which means that the potential grows linearly with distance $r$.
At finite temperature we can consider the spatial Wilson loops. They obey area law at any temperature $W_s(x,z) \sim \exp(-\sigma_s(T) x z)$ \cite{polonyi,bali93}.  
Below the transition temperature the spatial string tension is very close to the usual zero temperature string tension.
Well above the deconfinement transition temperature the spatial string tension is expected to be $\sqrt{\sigma_s(T)}=c_M g^2(T) T$ \cite{bali93}. This is because
in the dimensionally reduced theory it is given by $c_M g_3^2$ and at leading order the 3-dimensional gauge coupling is $g_3^2=g^2(T) T$. 
The spatial string tension has been calculated on the lattice in quenched QCD ($N_f=0$) \cite{lutge} and 2+1 flavor QCD \cite{liddle_lat06} and the results are shown in Figure \ref{fig:sp}.
The lattice data can be fitted very well  with the simple form : $\sqrt{\sigma_s(T)}=c_M g^2(T) T$.
Here again $g(T)$ is the 2-loop running coupling. For the fit  we get the value of $c_M$ which agrees reasonably well with the result of dimensional
reduction \cite{liddle_lat06}.  The 3-dimensional gauge coupling $g_3^2$ has been calculated more systematically in perturbation theory and also led to a very good
agreement with the lattice data \cite{laine,liddle_lat06}.
\begin{figure}
\rotatebox{-90}{\includegraphics[width=7cm]{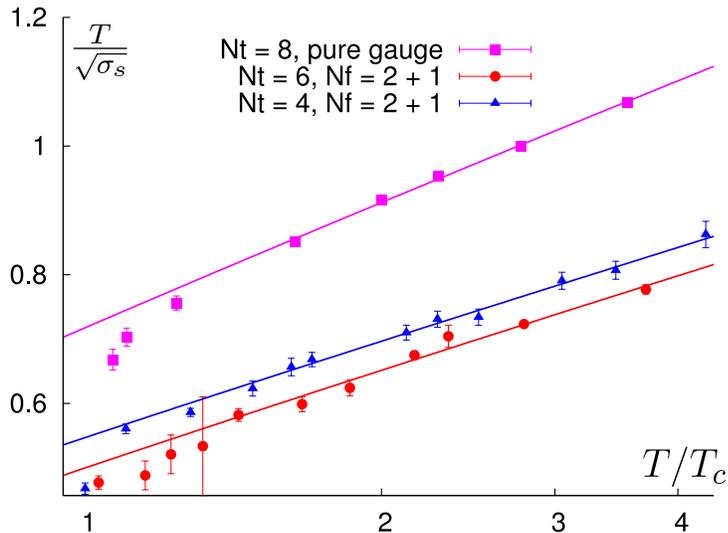}}
\caption{
The spatial string tension calculated in quenched QCD \cite{lutge} and 2+1 flavor QCD \cite{liddle_lat06}. The lines show the leading order
fits.}
\label{fig:sp}
\end{figure}

\section{Spectral functions}
Information on hadron properties at finite temperature as well as transport coefficients are encoded in different spectral functions.
In particular the fate of different quarkonium states in the quark gluon plasma can studied by calculating the corresponding quarkonium spectral functions.
On the lattice we can calculate correlation function in Euclidean time. The later is related to the spectral function via integral
relation
\begin{equation}
G(\tau, T) = \int_0^{\infty} d \omega
\sigma(\omega,T) K(\tau,\omega,T) ,~~
K(\tau,\omega,T) = \frac{\cosh(\omega(\tau-1/2
T))}{\sinh(\omega/2 T)}.
\label{eq.kernel}
\end{equation}
Given the data on the Euclidean meson correlator $G(\tau, T)$ the meson spectral function can be calculated
using the Maximum Entropy Method (MEM)  \cite{mem}. For charmonium this was done by using correlators calculated on
isotropic lattices \cite{datta02,datta04} as well as  anisotropic lattices \cite{umeda02,asakawa04,jako07} in quenched approximation.
It has been found that quarkonium correlation function in Euclidean time show only very small temperature
dependence \cite{datta04,jako07}. In other channels, namely the vector, scalar and axial-vector channel 
stronger temperature dependence was found \cite{datta04,jako07}, especially in the scalar and axial-vector channels.
The spectral functions in the pseudo-scalar and vector channels reconstructed from MEM show peak structures which may
be interpreted as a ground state peak \cite{umeda02,asakawa04,datta04}. Together with the weak temperature dependence
of the correlation functions this was taken as strong indication that the 1S charmonia ($\eta_c$ and $J/\psi$) survive
in the deconfined phase to temperatures as high as $1.6T_c$ \cite{umeda02,asakawa04,datta04}. A detailed study of
the systematic effects show, however, that the reconstruction of the charmonium spectral function is not reliable
at high temperatures \cite{jako07}. In particular the presence of peaks corresponding to bound states cannot be
reliably established. The only statement that can be made   is that the spectral function does not show significant changes 
within errors of the calculations. Recently quarkonium spectral functions have been studied using potential models
and lattice data for the free energy of static quark anti-quark pair \cite{mocsy07}. These calculations show that all
charmonia states are dissolved  at temperatures smaller than $1.5T_c$, but the Euclidean correlators do not show
significant changes and are in fairly good agreement with available lattice data both for charmonium \cite{datta04,jako07}
and bottomonium \cite{jako07,dattapanic05}. This is due to the fact that even in absence of bound states quarkonium spectral functions
show significant enhancement in the threshold region \cite{mocsy07}.  Therefore previous statements about quarkonia
survival at high temperatures have to be revisited. The large enhancement of the quarkonium correlators above deconfinement in the scalar and axial-vector
channel can be understood in terms of the zero mode contribution \cite{mocsy07,umeda07} and not due to the dissolution of
the $1P$ states as previously thought. Similar, though smaller in magnitude, enhancement of quarkonium correlators due to zero mode 
is seen also in the vector channel \cite{jako07}. Here it is related to heavy quark transport \cite{derek,mocsy06}.
Although the above mentioned lattice studies were performed in quenched
approximation we do not expect the picture to change when dynamical quarks are included in the calculations since
recent calculations in 2-flavor QCD show very similar temperature dependence of charmonium correlators \cite{aarts07}.

The spectral function for light mesons as well as the spectral function of the energy momentum tensor has been calculated on the lattice
in quenched approximation \cite{karsch02,asakawaqm02,aarts_el,meyer}. However, unlike in the quarkonium case 
the systematic errors in these calculations are not well understood. 

\section{Conclusions}
In recent years significant progress has been made in calculating bulk thermodynamic observables 
as well as spatial correlation functions on the lattice. These calculations suggest that at temperatures $T>2.0T_c$ thermodynamics
can be described reasonably well using weak coupling approaches: re-summed perturbation theory and dimensional
reduction. The temperature and flavor dependence of static screening length is well described by perturbation theory.
However, the value of the screening lengths is to large extent non-perturbative and influenced or determined by the
dynamics of static magnetic fields. Furthermore, there is no evidence for the large value of the gauge coupling constant at scale $T$.
Clearly more precise lattice data and further perturbative calculations are needed to establish the nature of quark gluon plasma
at $T>2.0T_c$. Despite significant progress our understanding of spectral functions at finite temperature is
still quite limited.

\section*{Acknowledgments}                                                                                                                         
This work has been supported by U.S. Department of Energy under Contract No. DE-AC02-98CH10886.

\vfill\eject
\end{document}